\documentclass[lettersize,journal]{IEEEtran}
\usepackage{amsmath,amsfonts}
\usepackage{algorithmic}
\usepackage{algorithm}
\usepackage{array}
\usepackage{textcomp}
\usepackage{stfloats}
\usepackage{url}
\usepackage{verbatim}
\usepackage{graphicx}
\usepackage{cite}
\usepackage{subcaption}
\usepackage{graphicx}
\usepackage{caption}
\usepackage{subcaption}
\usepackage{multirow}
\usepackage{xcolor}
\usepackage{dsfont}
\usepackage{tikz}
\usetikzlibrary{shapes.geometric, arrows,calc}
\usepackage{tikz}
\usepackage{pgfplots}

\begin{document}

\title{TeLeS: Temporal Lexeme Similarity Score to Estimate Confidence in End-to-End ASR}

\author{Nagarathna~Ravi*,~ Thishyan Raj T*,~ Vipul Arora
        % <-this % stops a space
\thanks{This paper was produced by the MADHAV Lab \texttt{ASR} research group in IIT Kanpur (e-mail: \{rathna, thishyan20, vipular\}@iitk.ac.in). *-Equal Contribution}% <-this % stops a space
\thanks{Manuscript received .... .., ....; revised .... .., .....}}

% The paper headers
\markboth{IEEE/ACM TRANSACTIONS ON AUDIO, SPEECH, AND LANGUAGE PROCESSING,~Vol.~.., No.~.., ....~..}%
{Ravi \MakeLowercase{\textit{et al.}}: Confidence Estimation for Low-Resource Automatic Speech Recognition}

%\IEEEpubid{0000--0000/00\$00.00~\copyright~2021 IEEE}
% Remember, if you use this you must call \IEEEpubidadjcol in the second
% column for its text to clear the IEEEpubid mark.

\maketitle

\begin{abstract}
% The abstract must be between 150-250 words.
Confidence estimation of predictions from an End-to-End (\texttt{E2E}) Automatic Speech Recognition (\texttt{ASR}) model benefits \texttt{ASR}'s downstream and upstream tasks. Class-probability-based confidence scores do not accurately represent the quality of overconfident \texttt{ASR} predictions. An ancillary Confidence Estimation Model (\texttt{CEM}) calibrates the predictions. State-of-the-art (\texttt{SOTA}) solutions use binary target scores for \texttt{CEM} training. However, the binary labels do not reveal the granular information of predicted words, such as temporal alignment between reference and hypothesis and whether the predicted word is entirely incorrect or contains spelling errors. Addressing this issue, we propose a novel \textbf{Te}mporal-\textbf{L}exeme \textbf{S}imilarity (\texttt{TeLeS}) confidence score to train \texttt{CEM}. To address the data imbalance of target scores while training \texttt{CEM}, we use shrinkage loss to focus on hard-to-learn data points and minimise the impact of easily learned data points. We conduct experiments with \texttt{ASR} models trained in three languages, namely Hindi, Tamil, and Kannada, with varying training data sizes. Experiments show that \texttt{TeLeS} generalises well across domains. To demonstrate the applicability of the proposed method, we formulate a TeLeS-based Acquisition (\texttt{TeLeS-A}) function for sampling uncertainty in active learning. We observe a significant reduction in the Word Error Rate (\texttt{WER}) as compared to \texttt{SOTA} methods.
\end{abstract}

\begin{IEEEkeywords}
Confidence Estimation, Automatic Speech Recognition, Human-in-the-Loop Annotation, Active Learning, Pseudo-labeling
\end{IEEEkeywords}

\section{Introduction}
\IEEEPARstart{C}{alibrating} an  End-to-End (\texttt{E2E}) Automatic Speech Recognition (\texttt{ASR}) model's predictions using confidence scores can aid downstream and upstream tasks \cite{i1}\cite{i1a}\cite{c13} such as domain adaptation, active learning, keyword spotting, dialogue systems, system combination, machine translation, information retrieval, semi-supervised learning, smart speakers, automatic error correction, dis-fluency detection, speaker adaptation, and enhancing the \texttt{ASR} dataset with diverse set of audio-transcript pairs.

\begin{figure}[ht]
    \centering
    \includegraphics[width=0.3\textwidth, height=4cm, keepaspectratio]{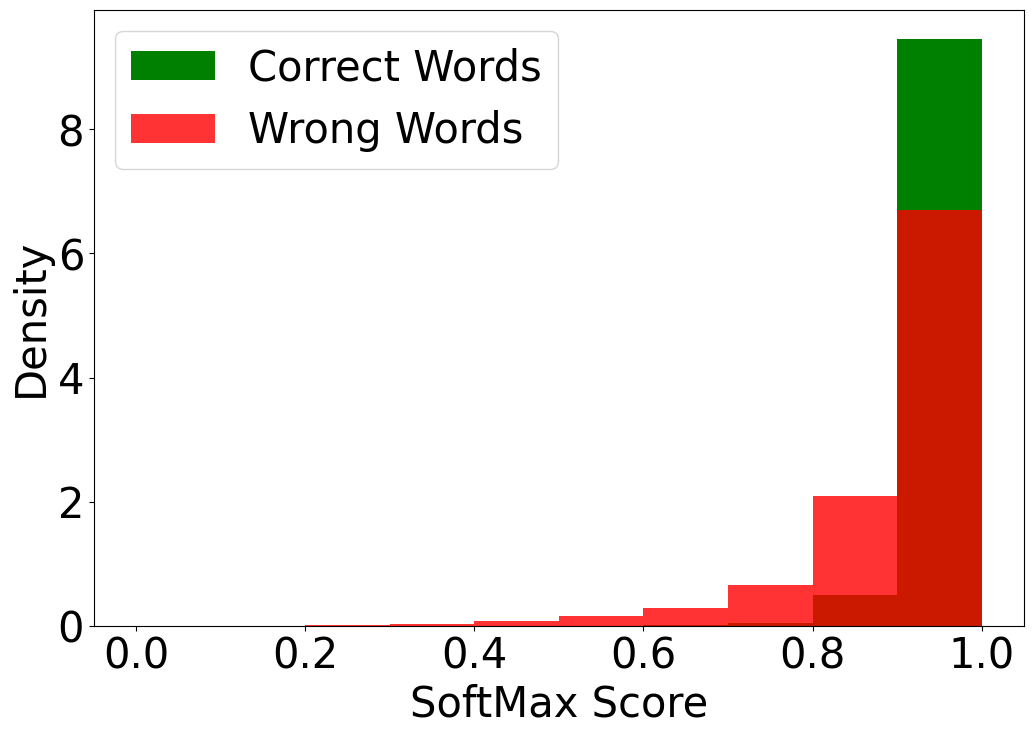}
    \caption{Class-prob. based Scores of Correct and Wrong Words}
    \label{fig1}
\end{figure}

The Confidence Estimation Model (\texttt{CEM}) can be an algorithm based on statistics \cite{i3} or an auxiliary Machine Learning (\texttt{ML}) model \cite{i2}. Statistical confidence measures include class-probability \cite{i4}\cite{i5} of the output label and entropy \cite{i3} of the class-probabilities. These measures assume that a well-learned model can accurately project the distance between the estimate and the label boundaries. However, as shown in Figure~\ref{fig1} (Class-probability-based confidence scores of \texttt{ASR} model trained using \texttt{IISc-Mile}-Tamil train dataset and tested on \texttt{IISc-Mile}-Tamil test-set \cite{iisc1}\cite{iisc2}\cite{iisc3}), there is a substantial overlap between the SoftMax scores of accurate and incorrect predictions due to the general overconfidence nature of neural models. Several State-Of-The-Art (\texttt{SOTA}) \cite{i2}\cite{i6} methods train an auxiliary \texttt{CEM} model using binary labels: score ’1’ is assigned to the correct label, and score ’0’ is set to the incorrect label as target values. Levenshtein alignment between the actual and hypothesis transcripts aids in identifying if the label is correct or wrong. However, using binary labels does not expose the predictions' granular information, such as the temporal alignment between the reference and the hypothesis. It also does not reveal whether the predicted word is entirely incorrect or contains spelling errors. Consider the example in which character alignments determine the target confidence scores of characters. When the input does not share the domain as the training data, alignments are ambiguous, as seen in the example (/ denotes the token boundary),\\
%[trim={left bottom right top},clip]
%/media/rathna/New Volume/taslp/misc.ipynb
%*स**ोना*** भी है
%बसा आना सच भी है
%['I', 'C', 'I', 'I', 'S', 'C', 'C', 'I', 'I', 'I', 'C', 'C', 'C', 'C', 'C', 'C']
Ref.: \includegraphics[scale=0.98,trim={0.2cm 0.15cm 0cm 0cm},clip]{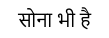}(/s/o/n/a/ /bh/ee/ /h/ai/)\\
Hyp.: \includegraphics[scale=0.98,trim={0.3cm 0.15cm 0 0},clip]{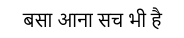}(/ba/s/a/ /aa/n/a/ /sa/ch/ /bh/ee/ /h/ai/)\\
Target Scores: $[0, 1, 0, 0, 0, 1, 1, 0, 0, 0, 1, 1, 1, 1, 1, 1]$\\
In the preceding example, the reference's first character \includegraphics[scale=0.4,trim={0.2cm 0.4cm 0cm 0cm},clip]{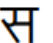} (s) is aligned with the second character \includegraphics[scale=0.4,trim={0.2cm 0.4cm 0cm 0cm},clip]{sa.pdf} (s) in the hypothesis. Similarly, the third character \includegraphics[scale=0.6,trim={0cm 0.4cm 0cm 0cm},clip]{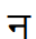} (n) is aligned with the sixth character \includegraphics[scale=0.6,trim={0cm 0.4cm 0cm 0cm},clip]{na.pdf} (n). However, the aligned characters are different, which shows that the method does not utilize temporal information. \\
Consider yet another example which assigns binary labels to words, \\
Ref.: \includegraphics[scale=0.98,trim={0cm 0.4cm 0.3cm 0}]{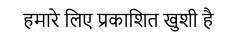} (/hamaare/ /lie/ /prakaashit/ /khushee/ /hai/)\\
Hyp.: \includegraphics[scale=0.98,trim={0.233cm 0.4cm 0.3cm 0}]{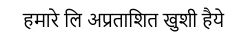} (/hamaare/ /li/ /aprataashit/ /khushee/ /haiye/)\\
Target Scores: $[1, 0, 0, 1, 0]$\\
The second word suffers one deletion. However, since the desired score is zero, the ancillary \texttt{CEM} attempts to penalize the estimation and push the score to zero. 

The need for confidence estimation in \texttt{ASR} for downstream and upstream tasks, issues with statistical confidence metrics, and insufficient information exposed by using binary labels as targets in training \texttt{CEM} motivate us to further probe into building a robust \texttt{CEM}.
This article presents a novel \texttt{CEM} mechanism to estimate the confidence of the words predicted by the \texttt{ASR} model. The mechanism involves training an auxiliary \textbf{W}ord-\textbf{L}evel \textbf{C}onfidence (\texttt{WLC}) model. The \texttt{WLC} model is trained with the intermediate outputs from the trained \texttt{ASR} model and a novel target confidence score $\in [0, 1]$. Training a robust \texttt{WLC} model requires balanced data points covering the target confidence score range. However, a well-trained \texttt{ASR} model generates limited words with errors, so the \texttt{WLC} model tends to be biased due to an imbalance in the dataset. \texttt{SOTA} methods like \cite{i6} do not have the methodology to handle an imbalance in the dataset. A few \texttt{SOTA} \cite{i2} methods introduce time and frequency masks in the audio spectrograms to introduce errors in the inference. However, it is uncertain whether getting data points across the target score range is possible. Hence, we propose to use the shrinkage loss \cite{shrinkage} function to train the \texttt{WLC} model. Shrinkage loss uses the sigmoid function to modulate the squared loss. Modulation aids in learning the infrequent samples and penalizes the significance of loss from frequent instances, which is our case's correct predicted words. We propose an acquisition function based on the estimated confidence scores from the \texttt{WLC} model for the active learning \texttt{ASR} downstream task to ascertain the usability of the proposed confidence model. The acquisition function chooses reliable pseudo labels and gives them to the annotators for correction, thereby creating a Human-In-The-Loop (\texttt{HITL}) architecture. The corrected transcripts are added to the training pool, ready for retraining the \texttt{ASR} model to adapt to the new data domain. In particular, the key contributions of our work are as follows.

\begin{itemize}
    \item We propose a novel confidence estimation metric called the \textbf{Te}mporal-\textbf{Le}xeme \textbf{S}imilarity score, abbreviated as \texttt{TeLeS}, to train the auxiliary \texttt{WLC} model. $TeLeS \in [0, 1]$. At frame level, the temporal similarity score reveals the misalignment between timestamps of the words in ground truth and hypothesis. The lexeme similarity score considers the number of deletions and insertions in the mismatched words in ground truth and inference. The combination of temporal and lexeme-level information aids in teaching the \texttt{TeLeS-WLC} model about the acoustic and decoder shifts that result in incorrect output. 
    \item To account for the data imbalance issue in training \texttt{TeLeS-WLC} model, we include shrinkage loss in the training architecture to aid the model in focusing on hard samples and avoid getting biased due to frequently cited samples. To the best of our knowledge, no works on confidence estimation have used shrinkage loss to account for the data imbalance issue.
    \item We propose the \texttt{TeLeS}-based Acquisition, \texttt{TeLeS-A} method to choose reliable samples for active learning-based \texttt{HITL} training. \texttt{TeLeS} aids in selecting informative samples for labelling and pseudo-labels for inclusion in the training loop.
    \item We provide an intensive evaluation of proposed methods in three different languages, viz., Hindi, Tamil, and Kannada. We evaluate the performance of the proposed methods utilizing a range of metrics, and the results are compared with \texttt{SOTA} methodologies. 
    \item We also release the Hindi dataset prepared by our in-house annotators as an open-source resource. Codes and sample audios of Hindi dataset are available online in \url{https://github.com/madhavlab/2023_teles_wlc} 
\end{itemize}

The remainder of the paper is structured as follows. Section II gives an overview of the existing methods in the literature that estimate the confidence of the outputs from the \texttt{ASR} model. Section III elaborates on the proposed methodology in detail. Section IV details the experimental design and outcomes gained while evaluating the \texttt{TeLeS}-\texttt{WLC} model. The results of using \texttt{TeLeS-A}-based active learning are presented in Section V. The final section concludes and gives future pointers for advancing the research further.

\section{Related Work}

\subsection{Confidence Estimation}
Confidence estimation in \texttt{ASR} has been actively explored for a few years. Class-probability score \cite{i4}\cite{i5} and its scaled version \cite{r1} are naive confidence estimation metrics. Entropy over prediction probabilities is also a statistical confidence measure \cite{r1}\cite{i3}. 

Huang et al. \cite{r5} propose training a deep neural network to predict whether or not the words or utterances are in-domain. Word or utterance identification data derived from the training corpus’s frequency statistics is used to train the network. However, a train data corpus cannot always represent real-time statistics.

Authors in \cite{r6} predict confidence using Recurrent Neural Networks (\texttt{RNN}). The acoustic and Language Models' (\texttt{LM}) input features are extracted and fed to \texttt{RNN} layers. Del-Agua et al. \cite{r7} train Deep Bidirectional \texttt{RNN} (\texttt{DBRNN}) using word-level predictor features and adapt the trained model to new speakers using a conservation strategy. Li et al. \cite{c13} propose estimating confidence using a \texttt{BRNN} that takes inputs like duration, embeddings, and word posterior probabilities from the \texttt{ASR} model. \cite{alexa} suggests extracting acoustic and hypothesis embedding to train a confidence estimator. The input Log Filter Bank Energy (\texttt{LFBE}) features are given to the Long Short Term Memory (\texttt{LSTM}) layer to get the acoustic embedding of each token. The authors extract each token's final \texttt{LSTM} decoder layer output as hypothesis embedding. The authors in \cite{c14} also employ a variant of the \texttt{RNN} to assess the accuracy of black-box \texttt{ASR} predictions. Jeon et al. \cite{c15} combine a heterogeneous word confusion network with a bidirectional lattice \texttt{RNN} to determine the most confident sequence. \cite{c16} proposes integrating \texttt{ELECTRA} with \texttt{ASR}. The proposed model leverages the notion of masked \texttt{LM} to learn the context and detect whether the output tokens from \texttt{ASR} are correct or have been substituted with another token. Ogawa et al. \cite{i6} extract multiple features from the \texttt{ASR} model, including the Connectionist Temporal Classification (\texttt{CTC}) score \cite{ctc}, attention vectors, \texttt{LM} scores, and the token's SoftMax probability score. A bidirectional-\texttt{LSTM} network gets the extracted features to estimate confidence. The authors use class-balanced cross-entropy loss to handle data imbalance in \texttt{CEM} train data. The method in \cite{c18} extracts acoustic and lexical information from the \texttt{ASR} model and parses them into a sequence layer to calibrate the predicted tokens. The word level score is calculated by averaging the token level scores. The proposed methodologies in \cite{i2}\cite{c20} extract various features from the \texttt{ASR} model, which include attention, embedding, SoftMax scores, and top-k SoftMax probabilities. The features are concatenated to serve as input to fully connected and residual-energy-based models in \cite{i2} and \cite{c20}, respectively. To assess the confidence of predictions from out-of-domain speech signals as well, the methodology in \cite{c21} integrates the \texttt{LM} with both methods described in \cite{i2}\cite{c20}. To handle the \texttt{CEM} data imbalance, \cite{i2}, \cite{c20}, and \cite{c21} use spectrogram augmention masking techniques to introduce errors in the \texttt{ASR} predictions during \texttt{CEM} training.  The approach in \cite{c22} employs a methodology similar to that used in \cite{i2}\cite{c20}. \texttt{CEM} in \cite{c22} collects features from the \texttt{ASR} model and calculates cross and self-attention of the extracted features. Attention vectors serve as inputs to the neural network for confidence estimation.

The techniques mentioned above take some features from the \texttt{ASR} model and parse them to auxiliary confidence estimation networks. Typically, the methods construct alignments at the token or word level between the ground truth transcript and the hypothesis. If the alignment between two words/tokens is correct, a score of '1' is assigned to the hypothesis index; otherwise, a score of '0' is assigned. The sequence of binary labels generated serves as the target confidence score for the auxiliary models. During training, the model attempts to reduce the gap between the output of the final layer and the target binary labels.

\subsection{\texttt{HITL}-Active Learning \texttt{ASR}}

This section discusses the \texttt{HITL} active learning approaches in the \texttt{ASR} domain. The \texttt{LM} Certainty (\texttt{LMC}) approach \cite{al1} uses \texttt{LM} perplexity to choose informative samples. The authors train an n-gram \texttt{LM} and feed the \texttt{ASR} predictions to the \texttt{LM} to obtain a perplexity score. Finally, the samples above a threshold are marked as informative data. Signal-Model Committee Approach (\texttt{SMCA}) \cite{al1} varies the dropout in the trained model and computes the character-matching error rate between the inferences generated by models with different dropouts. The dataset tuples with error rates above a threshold are marked as informative samples. Methods in \cite{al2} and \cite{al3} use length-normalized path probability of the \texttt{ASR} predictions as the active learning acquisition metric. The methodologies choose the samples with low path probability to annotate as they are more informative.

\subsection{Advantages over \texttt{SOTA}}

Next, we highlight some key advantages the proposed method provides over the \texttt{SOTA} methodologies.

\begin{itemize}
    \item \texttt{ASR} model attempts to overfit the data during training and tends to be overconfident in predicting unobserved data \cite{r3}\cite{r4}. So, the scores based on prediction probabilities in \cite{i4}, \cite{i5}, \cite{r1}, and \cite{i3} are an unreliable measure. The \texttt{TeLeS} score uses the ground truth to obtain the target scores for training \texttt{WLC} model; hence, the overconfident \texttt{ASR} predictions do not impact the confidence estimation.
    \item \cite{i2}, \cite{i6}, \cite{r6}-\cite{c22} use binary target scores. \texttt{TeLeS} score gives additional information when compared to binary targets. Hence, \texttt{TeLeS} target scores reveal the confidence of the predictions more effectively than the binary target scores.
    \item \cite{r6}, \cite{r7}, \cite{c13}, \cite{alexa}, \cite{c14}, \cite{c18} and \cite{c22} do not have methodologies to handle data imbalance in training \texttt{CEM}. Whereas, we use shrinkage loss that focuses on infrequent samples while training the \texttt{WLC} model.
    \item The active learning \texttt{LMC} approach in \cite{al1} uses \texttt{LM} trained on the source dataset. The \texttt{LMC} approach may not capture the informative samples if there is a domain mismatch. \texttt{SMCA} \cite{al1} and methods in \cite{al2} and \cite{al3} depend on the \texttt{ASR} prediction probabilities. Our approach uses estimated confidence, computed based on intermediate states from the \texttt{ASR} model and \texttt{TeLeS} target score.
\end{itemize}

\section{Proposed Methodology}

This section begins by introducing the task of speech recognition and estimating the confidence of recognition. %, and providing the necessary notations. 
Further, we introduce the confidence metric \texttt{TeLeS}. Then, we describe our \texttt{TeLeS-WLC} training approach that learns the confidence measures. Finally, we present the \texttt{TeLeS-A} function that can be utilised for active learning tasks in \texttt{ASR}.

\subsection{Problem Formulation}
%https://latex-programming.fandom.com/wiki/List_of_LaTeX_symbols - reference for math symbols
Consider the set of audio-transcript pairs $\mathcal{D} = \{(\mathbf{\tilde{x}}_i, \mathbf{z}_{i})\}_{i=1}^{K}$, where $\mathbf{\tilde{x}}_{i}$ is a raw speech audio file and $\mathbf{z}_{i}$ is the ground-truth transcript of $\mathbf{\tilde{x}}_{i}$. Speech-to-text conversion comprises taking $\mathbf{\tilde{x}}_i$ or its spectral representation and extracting from it basic sound units ($L$), such as phonemes, word-pieces, graphemes, characters and so on. Due to the quasi-stationary nature of speech signals, it is preferable to predict the sound units present in small overlapping time-frames, over which the characteristics of the speech signal remains stationary. In speech processing, MelSpectrogram feature representation is one of the most popular choices, as this represents the energies of the frequencies present in $\mathbf{\tilde{x}}_i$, in accordance with the Mel frequency scale. 

The MelSpectrogram representation of the audio signals and labels form the train dataset $\mathcal{D^\prime} = \{(\text{X}_i, \mathbf{z}_{i})\}_{i=1}^{K}$. From this point onward, we consider each $(\text{X},\mathbf{z}) \in \mathcal{D}^\prime$ as a general data point used for training our model. $\text{X} = \left[ \mathbf{x}_{1},\mathbf{x}_{2},\ldots, \mathbf{x}_{T} \right]$ with each $\mathbf{x}_{t} \in \mathbb{R}^{D}$ ($\mathbf{x}_{t}$ is the $D$-dimensional MelSpectrogram representation for the $t^\text{th}$ frame, $T$ is the number of frames for each $\text{X}$). There is a corresponding $\mathbf{z} = \left[ z_{1}, z_{2}, \ldots, z_{U} \right]$ which is the label sequence for $\text{X}$ and $U\le T$ is the length of the label sequence. Each $\mathbf{z}$ is from the set of all possible sequences made up of $L$ and each $z_{u} \in L$.

Models can be trained to classify the frames of $\text{X}$ as $z_u$. In our case, we train an Attention-based Deep Neural Network (\texttt{A-DNN}) model $\mathcal{F}$ with parameters $\theta$ using \texttt{CTC} loss \cite{ctc} to extract frame-wise context information and predict the utterance. $\mathcal{F}$ is a combination of two cascaded networks, namely, encoder $e$ and decoder $d$. 

%Given an input $s_n$, a trained $A-DNN$ $\mathcal{F}$ with parameters $\mathfrak{w}$ gives non-negative values for all token units in $\mathcal{U}$. The non-negative values are further normalized using softmax function to squash them to range 

%In our case, we train an Attention-based Deep Neural Network (\texttt{A-DNN}) model to extract frame-wise context information that is encoded into an attention matrix. 

%Consider the MelSpectrogram representation for an audio signal $\mathbf{X} = \left[\mathbf{x}_1, \mathbf{x}_2, \ldots, \mathbf{x}_t, \ldots, \mathbf{x}_T \right]$, where each $\mathbf{x}_t$ is the audio signal representation for the $t$th frame. 
The network $e$ extracts attentions from a sub-sampled version of the $\text{X}$ as $\text{A}= e(\text{X}) = [\mathbf{a}_1, \mathbf{a}_2, \ldots, \mathbf{a}_{T^\prime}]$ where $T^\prime$ is the number of frames after sub-sampling and $\mathbf{a}_{t^\prime}$ is the attention state vector at the $t^\prime$ sub-sampled frame. The reduction in number of time frames depends on the number of sub-sampling layers used in the attention network. If $k$ sub-sampling layers are used then $k T^\prime = T$, i.e., each frame of the sub-sampled representation contains information from $k$ frames of the original representation. Attention matrix $\text{A}$ is then given to a decoder network to obtain,
 \begin{align}
    \text{Y} , \text{H} & = d(\text{A})  \\
   \text{S} & = \text{SoftMax}(\text{Y})
\end{align}
where $\text{Y} = [\mathbf{y}_1, \mathbf{y}_2, \ldots, \mathbf{y}_{T^\prime}]$ is the output sequence of the decoder network, $\text{H} = [\mathbf{h}_1, \mathbf{h}_2, \ldots, \mathbf{h}_{T^\prime}]$ is the output of the decoder layer of the decoder network and $\text{S} = [\mathbf{s}_1, \mathbf{s}_2, \ldots, \mathbf{s}_{T^\prime}]$ are the frame-wise probability distributions over $L^\prime = L \cup \{blank\}$. We use $\text{S}$ to determine the frame-wise sound units by greedy decoding as, $\mathbf{\hat{z}} = \left[ \hat{z}_1, \hat{z}_2, \ldots, \hat{z}_{T^\prime}\right]$ where 
\begin{align}
    \hat{z}_{t^\prime} = \underset{L^\prime}{\arg \max}\ s_{t^\prime}
\end{align}

$\mathbf{\hat{z}}$ is transformed to words that make up the sentence as $\mathbf{\hat{w}} = \left[\hat{w}_1, \hat{w}_2 \ldots, \hat{w}_{N^\prime} \right]$, where each $\hat{w}_n = \left\{\hat{z}_i, \ldots, \hat{z}_j \right\}$ with $1\le i \le j \le T^\prime$, using a many-to-one map $\mathcal{B}$ required for models trained with CTC \cite{ctc}. $\mathcal{B}$ removes the blank tokens and repeated tokens in $\mathbf{\hat{z}}$.

To make the predictions from $\mathcal{F}$ more reliable, we define an auxiliary \texttt{WLC} neural network $\mathcal{K}_\mathcal{F}$ with parameters $\hat{\theta}$ to estimate confidence $c \in [0, 1]$ of elements in $\mathbf{\hat{w}}$. To satisfy our objective that $c$ should be correlated with level of correctness of words in $\mathbf{\hat{w}}$, we train $\mathcal{K}_\mathcal{F}$ using \texttt{TeLeS} score.

\subsection{\texttt{TeLeS} Score as a Confidence Measure}
%A good confidence measure must be a true estimate of the probability of occurrence of outputs with a particular confidence value. For a confidence measure $m$ to be a good estimate of the output of a model, 
%\begin{align}
%    p(c = \hat{c}_l | m=\hat{m}_l) = \hat{m}_l
%\end{align}
%Using softmax values to compute word-level scores as 
%\begin{align}
%    \hat{m}_l = \frac{1}{j-i+1} \sum_{l=i}^{j} p(\mathbf{y}_l | \mathbf{a}_l, \mathbf{y}_{1:l-1})
%\end{align} 
%for a word $\hat{w}_n$ is not a good confidence measure as models tend to be overconfident about bad predictions also. This is can be seen in figure \ref{fig1}. 

\texttt{TeLeS} score matches the temporal alignment of the predicted words with the temporal alignment of the reference words, and the similarity between the aligned reference and predicted words. 

We use Kaldi speech recognition toolkit to determine the temporal alignment of the words, $\mathbf{w}= \left[w_1, w_2,\ldots, w_{N} \right]$ in ground truth. The dataset $\mathcal{D^\prime}$ is used to train a Gaussian Mixture Model-Hidden Markov Model (\texttt{GMM-HMM}) based acoustic model. The trained \texttt{GMM-HMM} model is used to perform forced alignment to obtain the start and stop times of the $\mathbf{w}$, $\left[(w_1^{ST}, w_1^{ET}), (w_2^{ST}, w_2^{ET}), \ldots, (w_{N}^{ST}, w_{N}^{ET}) \right]$. 

We approximate the start and end times of words in $\mathbf{\hat{w}} = \left[(\hat{w}_1^{ST}, \hat{w}_1^{ET}), (\hat{w}_2^{ST}, \hat{w}_2^{ET}), \ldots, (\hat{w}_{N^\prime}^{ST}, \hat{w}_{N^\prime}^{ET}) \right]$ from the frame-level indices of the predicted sequence, $\mathbf{\hat{z}}$. 

We perform word level alignment between the reference and the prediction using Levenshtein's alignment to ensure that the temporal and lexical similarity scores are being computed between corresponding words. For the reference word sequence $\mathbf{w}$, and the predicted sequence $\mathbf{\hat{w}}$, we define the function \textsc{AliGen} to generate the Levenshtein alignments as,
\begin{align}
    G = \text{\textsc{AliGen}}(\mathbf{w}, \mathbf{\hat{w}})
\end{align}
where $G = [g_1, g_2, \ldots g_P]$ with $g_p \in \left\{ C,S,I,D\right\}$, which denotes correct, substitution, insertion and deletion respectively. We now define the temporal agreement score $c^T$ as 
\begin{align}
    c^T = \begin{cases}
        \max \left( 0, 1-\frac{\left|w_i^{ST} - \hat{w}_j^{ST} \right| + \left| w_i^{ET} - \hat{w}_j^{ET} \right|}{\left| w_i^{ET} - w_j^{ST}\right|} \right) & \text{if } g_p = C,S \\
        0 & \text{if } g_p = I, D
    \end{cases}
\end{align}
When the words are substituted, the words are not necessarily entirely wrong but they may be wrong due to spelling mistakes. We compare the lexical similarity score for these cases as it is essential to calibrate the model's confidence. We define the lexical similarity score $c^L$ as
\begin{align}
    c^L = \begin{cases}
        \frac{\left| w_i \cap \hat{w}_j \right|}{\left| w_i \cup \hat{w}_j\right|} & \text{if } g_p = C, S \\
        0 & \text{if } g_p = I, D
    \end{cases}
\end{align}
With these two components, we define the \texttt{TeLeS} score for every word as 
\begin{align}
    c = \begin{cases}
        \alpha \times c^L + (1-\alpha) \times c^T & \text{if } g_p = C \\
        \beta \times c^L + (1-\beta) \times c^T & \text{if } g_p = S \\
        0 & \text{if } g_p = I, D
    \end{cases}
\end{align}
where, $\alpha \in [0,1] $ and $\beta \in [0,1]$ are weights for temporal and lexeme similarity scores and they are tunable hyper-parameters. 

\subsection{Learning to Predict Confidence with \texttt{TeLeS-WLC}}

\begin{figure}[ht]
    \centering
    \includegraphics[width=0.5\textwidth, height=8cm, keepaspectratio]{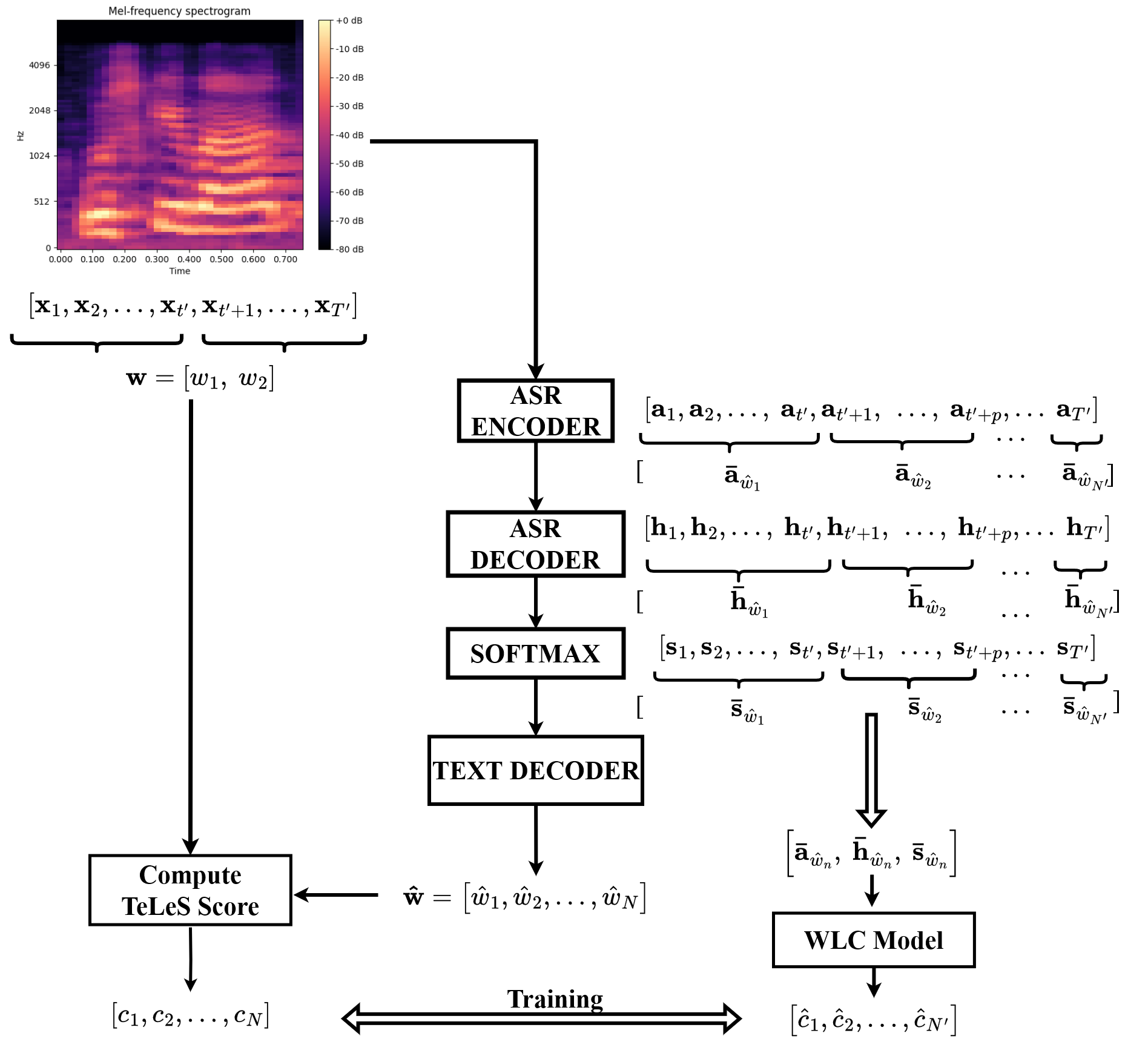}
    \caption{TeLeS-WLC Train Architecture}
    \label{arch}
\end{figure}

Figure~\ref{arch} gives an overview of the training architecture of \texttt{TeLes-WLC} model. \texttt{TeLeS-WLC} model is trained using intermediate states of \texttt{ASR} model and target \texttt{TeLeS} score. We fetch attention states $\text{A}$, decoder states $\text{H}$, and class probabilities $\text{S}$ from \texttt{ASR} model.

Using $\mathbf{\hat{z}}$ across $T^\prime$ frames, for each word $\hat{w}_n$, we find start and end indices of $\hat{w}_n$, $I_n = (b_{\hat{w}_n}, l_{\hat{w}_n})$, where $b_{\hat{w}_n}$ is the start index and $l_{\hat{w}_n}$ is the end index, satisfying the conditions $1 \leq b_{\hat{w}_n} \leq T^\prime$ and $b_{\hat{w}_n} \leq l_{\hat{w}_n} \leq T^\prime$. For each $\hat{w}_n$, we can now compute, 
\begin{align}
    \mathbf{\bar{a}}_{\hat{w}_n} & =  \text{mean}(\mathbf{a}_{b_{\hat{w}_n}}, \ldots, \mathbf{a}_{l_{\hat{w}_n}}) \\
    \mathbf{\bar{h}}_{\hat{w}_n} & =  \text{mean}(\mathbf{h}_{b_{\hat{w}_n}}, \ldots, \mathbf{h}_{l_{\hat{w}_n}}) \\
    \mathbf{\bar{s}}_{\hat{w}_n} & =  \text{mean}(\mathbf{s}_{b_{\hat{w}_n}}, \ldots, \mathbf{s}_{l_{\hat{w}_n}})
\end{align}

We train $\mathcal{K}_\mathcal{F}$ with the following pairs $([\mathbf{\bar{a}}_{\hat{w}_n}, \mathbf{\bar{h}}_{\hat{w}_n}, \mathbf{\bar{s}}_{\hat{w}_n}], c_{\hat{w}_n})$, where $c_{\hat{w}_n}$ is the \texttt{TeLeS} score of word $\hat{w}_n$. We use shrinkage loss \cite{shrinkage} to train the \texttt{WLC} model. Shrinkage loss is defined as,

\begin{align}
    \mathcal{L} = \left[\frac{\frac{1}{N^\prime} \sum_{\hat{w}_n \in \mathbf{\hat{w}} }\left( \hat{c}_{\hat{w}_n}- c_{\hat{w}_n}\right)^2 \cdot e^{{\hat{c}}_{\hat{w}_n}} }{1+e^{\gamma \cdot \left(\kappa-\frac{1}{N^\prime} \sum_{\hat{w}_n \in \mathbf{\hat{w}}}\left| \hat{c}_{\hat{w}_n}- c_{\hat{w}_n}\right|\right)}}\right] 
\end{align}
where, \\
$\hat{c}_{\hat{w}_n} = \mathcal{K}_\mathcal{F}([\mathbf{\bar{a}}_{\hat{w}_n}, \mathbf{\bar{h}}_{\hat{w}_n}, \mathbf{\bar{s}}_{\hat{w}_n}])$

$\gamma$ and $\kappa$ are hyper-parameters. $\kappa \in [0, 1]$ penalizes easy-to-learn data points by controlling the importance of frequently seen samples and $\gamma \in [0, \infty)$ controls the shrinkage rate of the $L2$ loss between $\hat{c}_{\hat{w}_n}$ and $c_{\hat{w}_n}$. $e^{{\hat{c}}_{\hat{w}_n}}$ is weight factor to accentuate the importance of hard-to-learn data point.

\subsection{\texttt{TeLeS-A}}
Consider an unlabeled set of audio $\text{U}=\{\text{U}_i\}_{i=1}^N$. The cost of manually labelling $\text{U}$ may be higher than the budget. Active learning entails discovering a set of uncertain samples within the budget and transcribing it with a human annotator. We propose an acquisition function \texttt{TeLeS-A} that uses confidence estimation from the \texttt{TeLeS-WLC} model. \texttt{TeLeS-A} acquires uncertain samples and samples that the \texttt{ASR} model is confident about. The uncertain samples are chosen within the labelling budget and given to the annotator. A certain sample set is selected with the help of a threshold $\delta$ and the audio-pseudolabel pairs are included in the train set.
\begin{align}
    \mathbf{\hat{w}} & = \mathcal{F}(\text{U}_i) \\
    \hat{c}_{\hat{w}_n} & = \mathcal{K}_\mathcal{F}([\mathbf{\bar{a}}_{\hat{w}_n}, \mathbf{\bar{h}}_{\hat{w}_n}, \mathbf{\bar{s}}_{\hat{w}_n}]) \\
     A_{\text{U}_i} & = \frac{\sum_{n=1}^{|\mathbf{\hat{w}}|} \hat{c}_{\hat{w}_n}}{|\mathbf{\hat{w}}|} \\
    \hat{\text{U}} &= sort([A_{\text{U}_i}]_{i=1}^{N})
\end{align}

The top samples from $\hat{\text{U}}$ within the labelling budget are picked, annotated and added to the training pool. The audio-pseudo-label pairs of the samples with $A_{\text{U}_i} \geq \delta$ are added to the train set.
\section{\texttt{CEM} Evaluation}

\subsection{Dataset}
The efficiency of the \texttt{TeLeS}-\texttt{WLC} model is evaluated using speech datasets of three different Indian languages: Hindi, Tamil, and Kannada. The following details the dataset used to train the base \texttt{ASR} and \texttt{TeLeS}-\texttt{WLC} models.

\begin{itemize}
    \item \textit{Hindi-} The dataset is a combination of speech data from different domains. The size of the dataset is $\approx 100$ hours. It has $\approx 17.6$ hours of speech data from the Common Voice Hindi (\texttt{CVH}) dataset  \cite{CVH-RArdila}. Our annotators made minor corrections to \texttt{CVH}, such as replacing numeric digits with the uttered text, transliterating English words into Hindi, inserting missing words, and correcting spelling errors. The rest of the speech data ($\approx 78.21$ hours) are taken from Prasar Bharati archive and manually annotated. The manually annotated datasets, collectively called the \texttt{PB} Hindi dataset, are speech recordings from different scenarios, including read news, the public addresses of eminent personalities, and recordings during programmes by various people. Hence, the dataset combines a few hours of recordings from diverse domains.

    \item \textit{Tamil-}  We use the benchmark \texttt{IISc-MILE} Tamil speech corpus, which comprises $\approx 150$ hours of train and test speech dataset \cite{iisc1}\cite{iisc2}.

    \item \textit{Kannada-} Benchmark \texttt{IISc-MILE} Kannada dataset \cite{iisc1}\cite{iisc2}\cite{iisc4} is used for the Kannada language, which comprises $\approx 350$ hours of training and test speech dataset. 
\end{itemize}

Verifying if the \texttt{TeLeS}-\texttt{WLC} model performs better in mismatched domain datasets is imperative. Hence, we use the benchmark KathBath (\texttt{KB}) \cite{kb-tamil}\cite{kb-tamil1} test-known dataset of the three languages to evaluate the respective trained \texttt{TeLeS}-\texttt{WLC} models.

\texttt{ASR} models are trained using speech conformer architecture \cite{conformer} and \texttt{CTC} loss. We train the Hindi \texttt{ASR} model with two Nvidia V100 16GB GPUs \cite{sanganak}. Four 40GB Nvidia A100 GPUs are used to train Tamil and Kannada \texttt{ASR} models \cite{siddhi}. The ``small" version of the speech conformer is configured in the encoder and decoder layers of the architecture, resulting in 10M parameters. If necessary, the speech audios are downsampled at 16000 Hz, and two channels are converted into one. We extract 80-dimensional Melspectrogram features from the processed speech audios and give them as input to the training module. AdamW optimizer with a learning rate 5e-4 and transformer scheduler \cite{transformer} are configured during the training phase. The output of the \texttt{ASR} models is space and \texttt{CTC} blank symbol added with the smallest grapheme unit of the languages: 64 Hindi characters, 51 Tamil characters, and 65 Kannada characters. Table~\ref{miss} gives the Word Error Rate (\texttt{WER}) and Character Error Rate (\texttt{CER}) observed on the \texttt{PB} Hindi, \texttt{IISc-Mile} Tamil and Kannada testsets using a greedy decoder and beam-search decoder with 3gram \texttt{KenLM} \cite{ken}. To confirm if the \texttt{KB} datasets are from a mismatched domain, we check the error rates of the inferences from the \texttt{ASR} model and present them in Table~\ref{miss1}. It is evident from Table~\ref{miss1} that the \texttt{KB} datasets have domain shift with respect to the trained \texttt{ASR} model. 

\begin{table}[ht]
%\scriptsize
    \centering
   %\small
    \caption{Accuracy of \texttt{ASR} models on Testsets}
    \begin{tabular}{|p{2.3cm}|p{1.1cm}|p{1.1cm}|p{1.1cm}|p{1.1cm}|}
    \hline
\multirow{2}{*}{\textbf{Testset}} &
  \multicolumn{2}{|c|}{\textbf{Without LM}} &
  \multicolumn{2}{|c|}{\textbf{With LM}}  \\

   % \hline
         & \textbf{WER(\%)} & \textbf{CER(\%)} &  \textbf{WER(\%)} & \textbf{CER(\%)} \\
         \hline
         PB- Hindi & 29.80 & 12.86 & 25.15 & 11.89\\
         \hline
         IISc-Mile- Tamil & 25.72 & 4.27 & 22.26 & 3.88\\
         \hline
         IISc-Mile- Kannada & 23.66 & 4.27 & 18.52 & 3.48 \\
         \hline 
    \end{tabular}
    \label{miss}
\end{table}

\begin{table}[ht]
%\scriptsize
    \centering
   %\small
    \caption{Accuracy of \texttt{ASR} models on \texttt{KB} Test-known sets}
    \begin{tabular}{|p{2.3cm}|p{1.1cm}|p{1.1cm}|p{1.1cm}|p{1.1cm}|}
    \hline
\multirow{2}{*}{\textbf{Testset}} &
  \multicolumn{2}{|c|}{\textbf{Without LM}} &
  \multicolumn{2}{|c|}{\textbf{With LM}}  \\

   % \hline
         & \textbf{WER(\%)} & \textbf{CER(\%)} &  \textbf{WER(\%)} & \textbf{CER(\%)} \\
         \hline
         KB-Hindi &  37.76 & 14.44 & 32.26 & 13.23 \\
         \hline
         KB-Tamil & 58.71 & 14.68 & 50.23 & 13.48\\
         \hline
         KB-Kannada & 56.67 & 15.76 & 48.48 & 14.50 \\
         \hline

    \end{tabular}
    \label{miss1}
\end{table}

We train the \texttt{TeLeS-WLC} architecture on Nvidia 3070Ti GPU. The \texttt{TeLeS-WLC} neural network comprises three fully connected layers and a sigmoid layer. The three layers have 512, 256 and 128 neurons, respectively. Activation function Rectified Linear Unit is added in between the neural layers. The model training is done with the Adam optimizer and a learning rate of 0.0001 for 50 epochs. We employ grid search over a range of $\left[0, 1\right]$ and run 10 epochs during each run to tune the \texttt{TeLeS} score hyperparameters $\alpha$ and $\beta$. We find that the ($\alpha, \beta$) = (0.75, 0.5) combination provides better results than other combinations and use them in subsequent studies.

\subsection{Baseline Methodologies}

We use three baseline approaches. The confidence score generated from \texttt{ASR} class probabilities is an approach \cite{i4}\cite{i5}. The class probabilities of the grapheme tokens are averaged to get the word-level confidence scores. The second baseline approach involves finding the Tsallis entropy with exponential normalization of the class probabilities of the tokens \cite{r1}\cite{i3}. The minimum function is the aggregation function to obtain the word-level confidence measure \cite{i3}. The final baseline approach \cite{i2}, \cite{i6}, \cite{r6}-\cite{c22} involves training the confidence neural network with binary target scores extracted from the alignment between ground truth and hypothesis using Levenshtein alignment.

\subsection{Evaluation Metrics}

Following standard evaluation metrics are used to evaluate \texttt{TeLeS}-\texttt{WLC} model performance: \\
\textbf{Mean Absolute Error} (\texttt{MAE} $\downarrow$)- The \texttt{MAE} $\in [0, 1]$ measures the disparity between the estimated score $\hat{y}_i$ and the actual score $y_i$ for the $i^{\text{th}}$ word. \texttt{MAE} is defined as 
\begin{align}
\text{MAE}(Y, \hat{Y}) = \frac{1}{N} \sum_{i=1}^{N} \left| y_i - \hat{y}_i\right| 
\end{align}
\textbf{Kullback-Leibler Divergence score} (\texttt{KLD} $\downarrow$)- \texttt{KLD} $\in [0, \infty] $ evaluates the difference between the estimated and actual distributions of scores. For $N$ words with the actual score as $y_i$ and estimated score as $\hat{y}_i$ for the $i^{\text{th}}$ word, \texttt{KLD} is defined as 
\begin{align}
D_{\text{KL}}(Y \parallel \hat{Y}) = \frac{1}{N} \sum_{i=1}^{N} \left[y_i \log \frac{y_i}{\hat{y}_i} + \left(1-y_i\right) \log \frac{1-y_i}{1-\hat{y}_i} \right]
\end{align}
\textbf{Jensen–Shannon divergence} (\texttt{JSD} $\downarrow$)- \texttt{JSD} $\in [0, 1]$ is a smoothed version of \texttt{KLD}. 
\begin{align}\text{JSD}(Y \parallel \hat{Y}) = \frac{\left( D_{\text{KL}}(Y \parallel M) + D_{\text{KL}}(\hat{Y} \parallel M) \right) }{2} 
\end{align}
where $M = \dfrac{Y+\hat{Y}}{2}$.\\
\textbf{Normalized Cross Entropy} (\texttt{NCE} $\uparrow$)- \texttt{NCE} $\in [-\infty, 1]$ \cite{nce}\cite{nce1} evaluates the correlation between the estimated confidence score and base \texttt{ASR} model accuracy. 
\begin{align}
    H(p) & = - p\log p - (1-p) log (1-p) \\
    H(p,\hat{Y}) & = \frac{-1}{N}\left[ \sum_{i|g_i=C}  \log \hat{y}_i\ +\ \sum_{i|g_i\ne C} \log (1-\hat{y}_i) \right] \\
    \text{NCE} & = \frac{H(p) - H(p, \hat{Y})}{H(p)} 
\end{align}\\where $p$ is the accuracy of the \texttt{ASR} model and $g_i$ indicates the word status obtained from the Levenshtein alignment described in section III.B.\\
\textbf{Calibration Error} (\texttt{CE} $\downarrow$) The purpose of using confidence scores is to minimise the worst-case gap between estimated confidence and measured word accuracy \cite{ce}. The computation of \texttt{CE} requires binning. Expected Calibration Error (\texttt{ECE}) $\in [0, 1]$ and Maximum Calibration Error (\texttt{MCE}) $\in [0, 1]$ are two distinct \texttt{CE} metrics. \texttt{ECE} is the weighted mean of all confidence and accuracy gaps. In contrast, \texttt{MCE} is the most significant gap among all bins. 
\begin{align}
    \text{ECE} & = \sum_{m=1}^{M} \frac{B_m}{n} \left| acc(B_m) - conf(B_m) \right| \\
    \text{MCE} & = \underset{m\in {1,\ldots,M}}{\max} \left| acc(B_m) - conf(B_m)\right|
\end{align}
where $B_m$s are equally spaced `$M$' bins, $n$ is the total number of words, $conf(B_m) = \frac{1}{m_n} \sum_{n=1}^{m_n} \hat{y}_n$ for $ \hat{y}_n \in B_m$ and $acc(B_m) = \frac{1}{|B_m|} \sum_{i\in B_m} \mathds{1}(y_i = \hat{y}_i)$.\\
\textbf{Root Mean Square Error- Word Correctness Ratio} (\texttt{RMSE-WCR} $\downarrow$)- \texttt{RMSE-WCR} $\in [0, 1]$ evaluates the accuracy of the estimated confidence with respect to \texttt{WCR} \cite{c22}. 
\begin{align}
    \text{A}_i & = \frac{1}{N^\prime} \sum_{j=1}^{N^\prime} \hat{y}_j \\
    \text{WCR}_i & = \frac{\text{Number of correctly predicted words}}{\text{Total number of predicted words}} \\
    \text{RMSE WCR} & = \sqrt{\frac{1}{K^\prime}\sum_{i=1}^{K^\prime} (\text{A}_i - \text{WCR}_i)^2}
\end{align}

\texttt{NCE} and \texttt{CE} are not ideal evaluation metrics for our case. They require the target labels to be binary. \texttt{SOTA} \cite{i2}, \cite{i6}, \cite{r6}-\cite{c22} methods use binary target labels, and hence, their evaluation metrics include \texttt{NCE} and \texttt{CE}. For comparison, we compute \texttt{NCE} and \texttt{CE} between the predicted scores from \texttt{TeLeS-WLC} model and the binary labels from Levenshtein alignment. Hence, better measures for comparing continuous \texttt{TeLeS} targets and estimated scores are \texttt{MAE}, \texttt{KLD}, \texttt{JSD} and \texttt{RMSE-WCR}.

\subsection{Training Modes}

Our first step of the experiment is to verify which training mode yields better performance. Four different training combinations are coded as follows.
\begin{itemize}
    \item Train with \texttt{MAE} loss (\texttt{MAE-L}).
    \item Transform the audio on-the-fly by applying time and frequency masking on the respective spectrograms \cite{i2}\cite{c20}\cite{c21},  before generating \texttt{ASR} inferences and employ \texttt{MAE} loss for training (\texttt{Mask-MAE-L}).
    \item Train with shrinkage loss (\texttt{Shrink-L}).
    \item Apply time and frequency masking on-the-fly on the audio spectrograms before generating \texttt{ASR} inferences and use shrinkage loss for training (\texttt{Mask-Shrink-L}).
\end{itemize}

Spectrogram augmentation involves transformation using ten time masks with mask lengths of 20 and 0.05 as the time-steps proportion that can be masked, followed by frequency mask with a mask length of 27. The shrinkage loss parameters are configured as $\gamma = 5$, $\kappa = 0.2$, which we identify through grid search over ranges $[2, 20]$ for $\gamma$ and $[0.1, 1.0]$ for $\kappa$.

\begin{figure}[ht]
    \centering
    \includegraphics[width=0.3\textwidth, height=4cm, keepaspectratio]{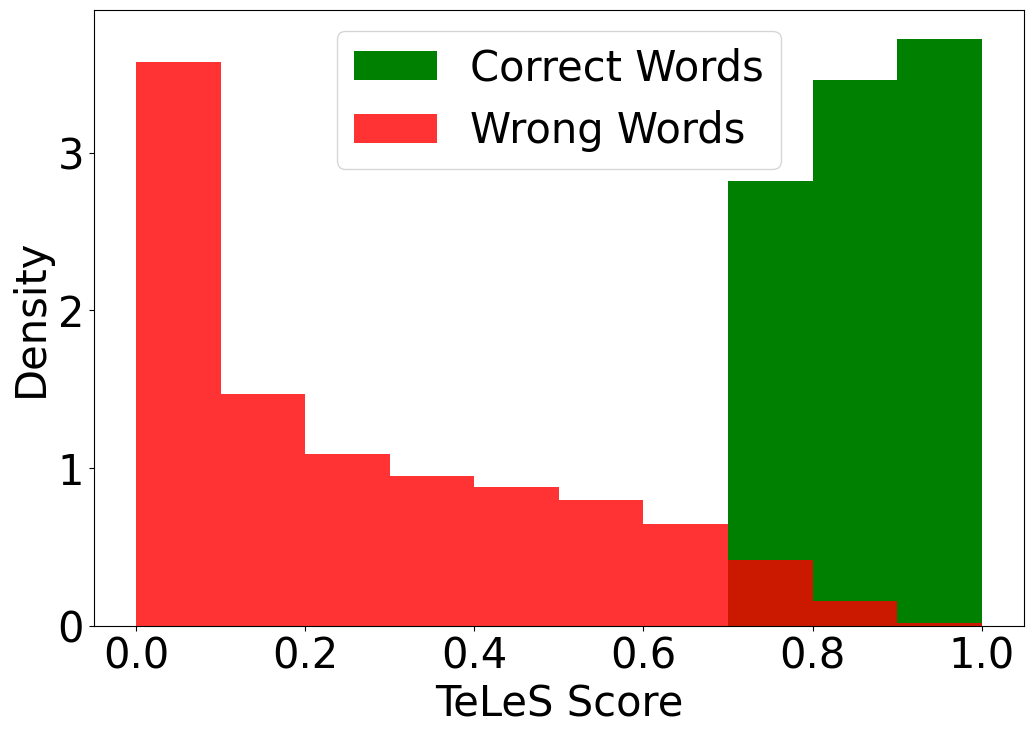}
    \caption{Distribution of Words across \texttt{TeLeS} Score Range}
    \label{balance}
\end{figure}

Table~\ref{loss} shows the results observed in the Hindi \texttt{ASR} (\texttt{PB} Hindi test) case for different training modes.  It is evident from the \texttt{ASR} model's \texttt{WER} in Table~\ref{miss} that the \texttt{WLC} model training sees a significant percentage of correct words. Hence, \texttt{MAE-L} and \texttt{Shrink-L} are influenced by correct word scores. To boost the rate of incorrect words, \cite{i2}\cite{c20}\cite{c21} transform the audio with time and frequency mask techniques. However, it does not produce a balanced set of words across the $[0, 1]$ \texttt{TeLes} score range, as evident from Figure~\ref{balance} (Score distribution taken from \texttt{PB} dataset). We can find a slight overlap between scores of correct and incorrect words. It is because some predicted words have minor spelling mistakes and temporal alignment with the ground truth. For instance, the words \includegraphics[scale=0.4,trim={0cm 0.2cm 0cm 0cm},clip]{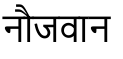} (/n/au/ja/v/aa/n/) and \includegraphics[scale=0.4,trim={0cm 0.4cm 0cm 0cm},clip]{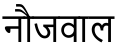} (/n/au/j/a/v/aa/l/) by one character and are temporally aligned with each other. Hence, we observe its \texttt{TeLeS} score as $0.816$.

\begin{figure}[ht]
    \centering
    \includegraphics[width=0.3\textwidth, height=4cm, keepaspectratio]{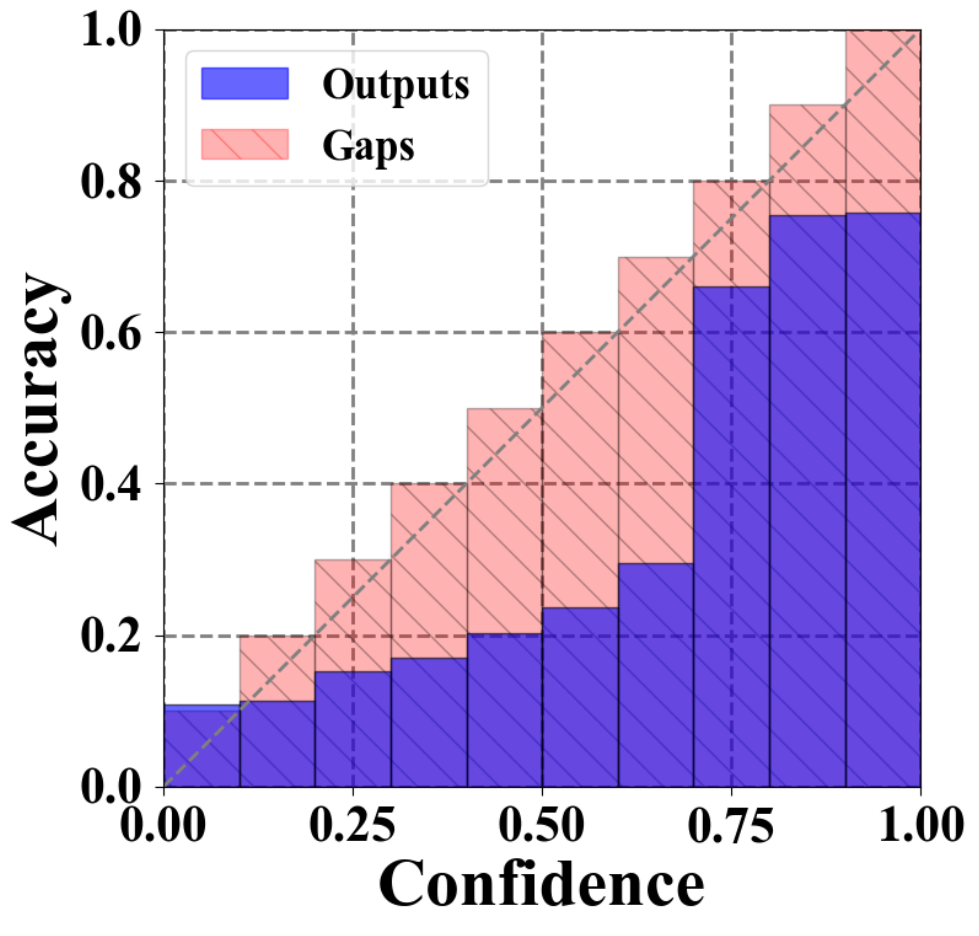}
    \caption{Calibration Curve - \texttt{Mask-MAE-L}}
    \label{maespec}
\end{figure}

\begin{figure}[ht]
    \centering
    \includegraphics[width=0.3\textwidth, height=4cm, keepaspectratio]{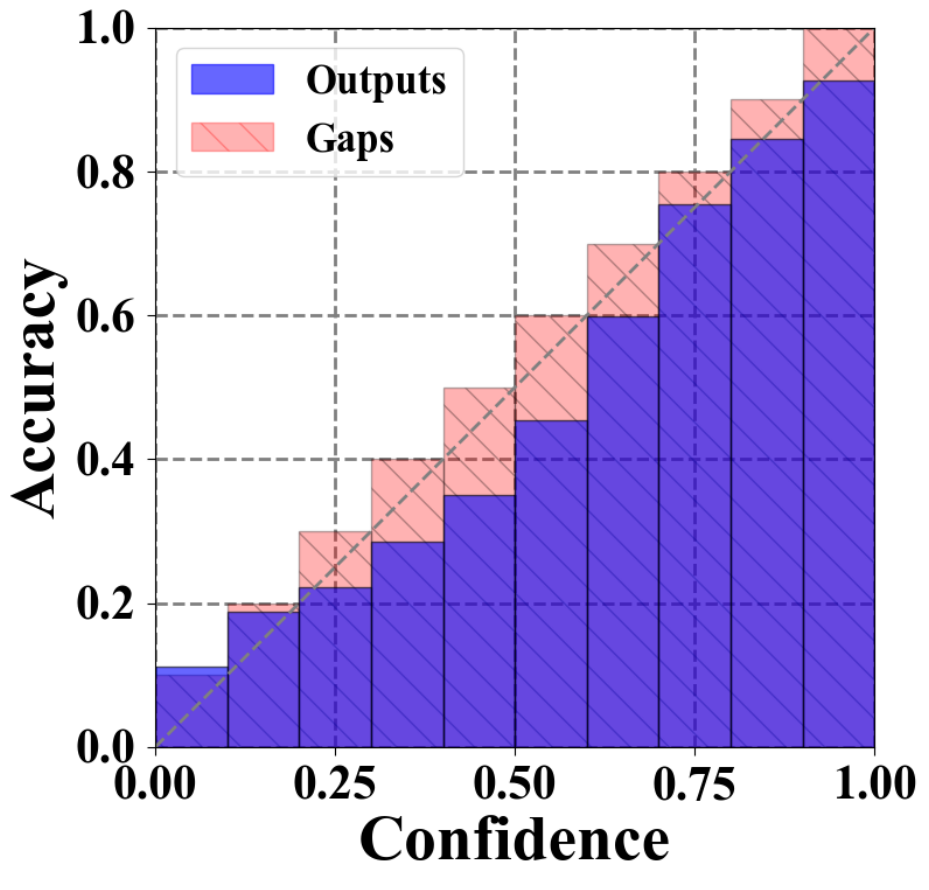}
    \caption{Calibration Curve - \texttt{Mask-Shrink-L}}
    \label{shrinkspec}
\end{figure}

\texttt{Mask-Shrink-L} train combination yields better results, as seen in Table~\ref{loss}. \texttt{MAE-L} has better \texttt{MAE} than the \texttt{Mask-Shrink-L}. However, other metrics show better results in \texttt{Mask-Shrink-L} case, as shrinkage loss handles the score imbalance. \texttt{CE} curves in Figures~\ref{maespec} and \ref{shrinkspec} also pictorially depict the efficacy of the same. Hence, in further experiments, we use \texttt{Mask-Shrink-L} for training.

\begin{table}[ht]
%\scriptsize
    \centering
   %\small
    \caption{Evaluation of \texttt{TeLeS-WLC} Training modes - Hindi \texttt{ASR} model}
    \begin{tabular}{|p{1.7cm}|p{1.2cm}|p{1.2cm}|p{1.2cm}|p{1.2cm}|}
    \hline
         \textbf{Metrics} & \textbf{MAE-L} & \textbf{Mask-MAE-L} & \textbf{Shrink-L} & \textbf{Mask-Shrink-L}\\
         \hline
         \texttt{MAE $\downarrow$}  & 0.2043 & \textbf{0.1783} & 0.1966  & 0.1817 \\
         \hline
        \texttt{KLD $\downarrow$} & 0.3680 & 0.2583 & 0.2809 & \textbf{0.1785} \\
         \hline
         \texttt{JSD $\downarrow$} & 0.0783 & 0.0595 & 0.0659  & \textbf{0.0467} \\
         \hline
         \texttt{RMSE-WCR $\downarrow$} & 0.2557 & 0.1987 & 0.2229 & \textbf{0.1457} \\
         \hline
          \texttt{NCE $\uparrow$} & -0.2948 & -0.0238 & -0.1272 & \textbf{0.1363} \\
         \hline
         \texttt{ECE $\downarrow$} & 0.1908 & 0.1345 & 0.1483 & \textbf{0.0260} \\
        
         \hline
         \texttt{MCE $\downarrow$} & 0.4265 & 0.3575 & 0.2671 & \textbf{0.1011} \\
         \hline

    \end{tabular}
    \label{loss}
\end{table}

\subsection{Ablation}

To verify if the combination of lexeme and temporal similarity is required for computing the \texttt{TeLeS} score, we study the performance of models trained using only lexeme similarity (i.e., $\alpha = 1$; $\beta = 1$) and temporal similarity ($\alpha = 0$; $\beta = 0$). Table~\ref{ablation} gives the results of the ablation study done with the Tamil language dataset (\texttt{IISc-Mile} Tamil testset). We see that \texttt{MAE} is better with lexeme-similarity only case. It can be partly due to the training of the model, wherein the training procedure tries to bring down the difference between predictions and targets. However, \texttt{RMSE-WCR} and \texttt{NCE} metrics that correlate the estimated confidence values with \texttt{ASR} model accuracy show that combining lexeme and temporal similarity via \texttt{TeLeS} score is more effective than using either.

\begin{table}[ht]
%\scriptsize
    \centering
   %\small
    \caption{Ablation Study on Components of \texttt{TeLeS} - Tamil}
    \begin{tabular}{|p{1.7cm}|p{1.7cm}|p{1.8cm}|p{1.7cm}|}
    \hline
         \textbf{Metrics} & \textbf{Lexeme only} & \textbf{Temporal only}  & \textbf{TeLeS}\\
         \hline
         \texttt{MAE $\downarrow$}  & \textbf{0.1296} & 0.1917  & 0.1489 \\
         \hline
        \texttt{KLD $\downarrow$} & 0.1942 &  0.2091 & \textbf{0.1432} \\
         \hline
         \texttt{JSD $\downarrow$} & 0.0596 &  0.0584 & \textbf{0.0373} \\
         \hline
         \texttt{RMSE-WCR $\downarrow$} & 0.2258 & 0.2405  & \textbf{0.1955}  \\
         \hline
          \texttt{NCE $\uparrow$} & -0.1277 &  -0.1741 & \textbf{0.0971} \\
         \hline
         \texttt{ECE $\downarrow$} & 0.1244 & 0.1312  & \textbf{0.0456} \\
        
         \hline
         \texttt{MCE $\downarrow$} & 0.2146 & 0.6528  & \textbf{0.1514} \\
         \hline
    \end{tabular}
    \label{ablation}
\end{table}

\subsection{Comparison with \texttt{SOTA}}

Tables~\ref{SOTA-hindi}, \ref{SOTA-tamil}, and \ref{SOTA-kannada} compare the results of various evaluation metrics observed during the experiment on the three languages with \texttt{SOTA} methods (\texttt{PB} and \texttt{IISc-Mile} test sets). Using class-probability-based \cite{i4}\cite{i5} and entropy-based \cite{r1}\cite{i3} confidence scores showcases degraded performance resulting from the overconfidence behaviour of \texttt{ASR}. Overconfidence behaviour fails to delineate a boundary between the scores of correct and incorrect words. \texttt{TeLeS}-\texttt{WLC} method shows better performance than using binary targets \cite{i2}, \cite{i6}, \cite{r6}-\cite{c22}. Binary targets try to penalize the predictions of words with minor mistakes and bring them down to 0. However, the input features of the words with minor errors will be close to those of the correct words. Hence, the model with binary targets could not learn the subtle variations in the patterns of the input features.

\begin{table}[ht]
%\scriptsize
    \centering
   %\small
    \caption{\texttt{TeLeS-WLC} and \texttt{SOTA} - Hindi \texttt{PB} Testset}
    \begin{tabular}{|p{1.7cm}|p{1.4cm}|p{1.2cm}|p{1.2cm}|p{1.1cm}|}
    \hline
         \textbf{Metrics} & \textbf{Class-Prob} & \textbf{Entropy}  & \textbf{Binary}  & \textbf{TeLeS}\\
         \hline
         \texttt{MAE $\downarrow$}  & 0.5026 & 0.3998 & 0.2650  & \textbf{0.1817} \\
         \hline
        \texttt{KLD $\downarrow$} & 0.7912 & 0.6333 & 1.1799  & \textbf{0.1785} \\
         \hline
         \texttt{JSD $\downarrow$} & 0.2313 & 0.1818 & 0.1920  & \textbf{0.0467} \\
         \hline
         \texttt{RMSE-WCR $\downarrow$} & 0.2883 & 0.2684 & 0.2593  & \textbf{0.1457}  \\
         \hline
          \texttt{NCE $\uparrow$} & -0.2641 & -0.0100  &  -0.0055 & \textbf{0.1363} \\
         \hline
         \texttt{ECE $\downarrow$} & 0.2472 & 0.2253 & 0.2627  & \textbf{0.0260} \\
        
         \hline
         \texttt{MCE $\downarrow$} & 0.3904 & 0.3663  & 0.4294  & \textbf{0.1011} \\
         \hline

    \end{tabular}
    \label{SOTA-hindi}
\end{table}

\begin{table}[ht]
%\scriptsize
    \centering
   %\small
    \caption{\texttt{TeLeS-WLC} and \texttt{SOTA} - Tamil \texttt{IISc-MILE} Testset}
    \begin{tabular}{|p{1.7cm}|p{1.4cm}|p{1.2cm}|p{1.2cm}|p{1.1cm}|}
    \hline
         \textbf{Metrics} & \textbf{Class-Prob} & \textbf{Entropy} & \textbf{Binary} & \textbf{TeLeS}\\
         \hline
         \texttt{MAE $\downarrow$}  & 0.5957 & 0.5618 & 0.2158  & \textbf{0.1489} \\
         \hline
        \texttt{KLD $\downarrow$} & 1.0214 & 0.9849 & 0.9579  & \textbf{0.1432} \\
         \hline
         \texttt{JSD $\downarrow$} & 0.2871 & 0.2726 & 0.1544  & \textbf{0.0373} \\
         \hline
         \texttt{RMSE-WCR $\downarrow$} & 0.4993 & 0.4956 & 0.2441  & \textbf{0.1955}  \\
         \hline
          \texttt{NCE $\uparrow$} & -0.9016 & -0.8337 &  -0.2912 & \textbf{0.0971} \\
         \hline
         \texttt{ECE $\downarrow$} & 0.4495 & 0.4579 & 0.2084  & \textbf{0.0456} \\
        
         \hline
         \texttt{MCE $\downarrow$} & 0.8958 & 0.5437 & 0.4330  & \textbf{0.1514} \\
         \hline

    \end{tabular}
    \label{SOTA-tamil}
\end{table}

\begin{table}[ht]
%\scriptsize
    \centering
   %\small
    \caption{\texttt{TeLeS-WLC} and \texttt{SOTA} - Kannada \texttt{IISc-MILE} Testset}
   \begin{tabular}{|p{1.7cm}|p{1.4cm}|p{1.2cm}|p{1.2cm}|p{1.1cm}|}
    \hline
        \textbf{Metrics} & \textbf{Class-Prob} & \textbf{Entropy} & \textbf{Binary} & \textbf{TeLeS}\\
         \hline
         \texttt{MAE $\downarrow$}  & 0.5350 & 0.6539 &  0.1845 & \textbf{0.1261} \\
         \hline
        \texttt{KLD $\downarrow$} & 0.8632  & 1.4422 & 0.8297  & \textbf{0.0927} \\
         \hline
         \texttt{JSD $\downarrow$} & 0.2502  & 0.3518 & 0.1332  & \textbf{0.0246} \\
         \hline
         \texttt{RMSE-WCR $\downarrow$} & 0.3946 & 0.6289 & 0.1780  & \textbf{0.1413} \\
         \hline
          \texttt{NCE $\uparrow$} & -0.6703 & -1.7907 & -0.4396  &  \textbf{0.1584}\\
         \hline
         \texttt{ECE $\downarrow$} & 0.3537 & 0.6126 &  0.1821 & \textbf{0.0699} \\
        
         \hline
         \texttt{MCE $\downarrow$} & 0.6122 & 0.6935 &  0.3597 & \textbf{0.1100} \\
         \hline 
         
    \end{tabular}
    \label{SOTA-kannada}
\end{table}

\subsection{Evaluation of \texttt{TeLeS-WLC} on Mismatched Domain}

The \texttt{TeLeS}-\texttt{WLC} model shows better results for in-domain speech data. Table~\ref{mis} shows the performance of the \texttt{TeLeS}-\texttt{WLC} model on the mismatched \texttt{KB} test-known datasets for Hindi, Tamil and Kannada. We find that the performance is better for all three languages, which indicates that \texttt{TeLeS} generalizes well. Table~\ref{refpred} gives examples of reference, hypothesis and confidence scores given by \texttt{TeLeS-WLC} (Rough transliteration is given).

\begin{table*}[ht]
%\scriptsize
    \centering
   %\small
    \caption{Estimated TeLeS Confidence Score}
   \begin{tabular}{|p{1.2cm}|p{5.1cm}|p{5.1cm}|p{2cm}|p{2.6cm}|}
    \hline
        \textbf{Language} & \textbf{Reference} & \textbf{Prediction} & \textbf{Alignment} & \textbf{Estimated TeLeS}\\
         \hline
         Hindi  & \includegraphics[scale=0.7,trim={0.2cm 0.4cm 0.3cm 0}]{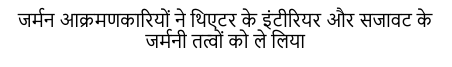} (/jarman/ /aakramanakaariyon/ /ne/ /thietar/ /ke/ /inteeriyar/ /aur/ /sajaavat/ /ke/ /jarmanee/ /tatvon/ /ko/ /le/ /liya/) & \includegraphics[scale=0.7,trim={0.1cm 0.6cm 0.3cm 0}]{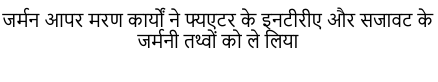}  (/jarman/ /aapar/ /maran/ /kaaryon/ /ne/ /phyetar/ /ke/ /inateereee/ /aur/ /sajaavat/ /ke/ /jarmanee/ /tathvon/ /ko/ /le/ /liya/) &  ['C', 'I', 'I', 'S', 'C', 'S', 'C', 'S', 'C', 'C', 'C', 'C', 'S', 'C', 'C', 'C'] & [0.67, 0.37, 0.43, 0.42, 0.81, 0.50, 0.78, 0.18, 0.69, 0.82, 0.81, 0.77, 0.42, 0.86, 0.72, 0.69] \\
         \hline
        Tamil & \includegraphics[scale=0.6,trim={0cm 0.4cm 0.3cm 0}]{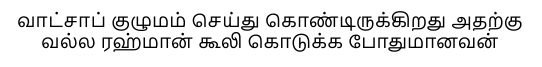} (/whatsapp/ /kuzhumam/ /ceydhu/ /kondirukkiradhu/ /adharku/ /valla/ /rahman/ /kooli/ /kodukka/ /podhumaanavan/)
 & \includegraphics[scale=0.6,trim={0cm 0.4cm 0.3cm 0}]{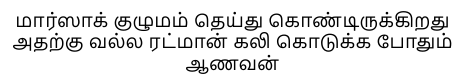} (/Maarshak/ /kuzhumam/ /teydhu/ /kondirukkiradhu/ /adharku/ /valla/ /ratman/ /kali/ /kodukka/ /podhum/ /aanavan/) 
 & ['S', 'C', 'S', 'C', 'C', 'C', 'S', 'S', 'C', 'I', 'S']  & [0.48, 0.82, 0.51, 0.86, 0.81, 0.74, 0.58, 0.62, 0.75, 0.52, 0.53] \\
         \hline
         Kannada & \includegraphics[scale=0.7,trim={0cm 0.4cm 0.3cm 0}]{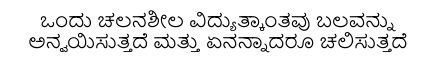} 
(/Ondhu/ /chalanasheela/ /vidyutkaantavu/ /balavannu/ /anvayisutthadhey/ /matthu/ /enannaadaru/ /chalisutthadey/)  & \includegraphics[scale=0.7,trim={0cm 0.4cm 0.3cm 0}]{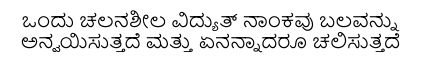} 
(/Ondu/ /chalanasheela/ /vidyut/ /naankavu/ /balavannu/ /anvayisutthadhey/ /matthu/ /enannadaru/ /chalisutthadey/) & ['C', 'C', 'I', 'S', 'C', 'C', 'C', 'C', 'C']  & [0.77, 0.73, 0.55, 0.32, 0.83, 0.77, 0.78, 0.79, 0.70] \\
         \hline

    \end{tabular}
    \label{refpred}
\end{table*}

\begin{table}[ht]
%\scriptsize
    \centering
   %\small
    \caption{Performance of \texttt{TeLeS-WLC} on KB Test-Known Datasets}
    \begin{tabular}{|p{1.7cm}|p{1.3cm}|p{1.3cm}|p{1.3cm}|}
    \hline
         \textbf{Metrics} & \textbf{Hindi} & \textbf{Tamil} & \textbf{Kannada} \\
         \hline
         \texttt{MAE $\downarrow$}  & 0.1869 & 0.2280 &  0.1925  \\
         \hline
        \texttt{KLD $\downarrow$} & 0.1804 & 0.2217 &  0.1621  \\
         \hline
         \texttt{JSD $\downarrow$} & 0.0478 & 0.0588  &  0.0437  \\
         \hline
         \texttt{RMSE-WCR $\downarrow$} & 0.1670 & 0.2700 & 0.2181   \\
         \hline
          \texttt{NCE $\uparrow$} & 0.1590 & 0.1151 & 0.2156   \\
         \hline
         \texttt{ECE $\downarrow$} & 0.0593 & 0.2116 &  0.1460  \\
        
         \hline
         \texttt{MCE $\downarrow$} & 0.1552 & 0.2643  & 0.2745   \\
         \hline
         
    \end{tabular}
    \label{mis}
\end{table}

\section{\texttt{TeLeS-A} Evaluation}

\subsection{Data and Baseline}

For active learning experiments, we use the base \texttt{ASR} models trained with \texttt{PB}-Hindi, \texttt{IISc-MILE} Tamil and \texttt{IISc-MILE} Kannada datasets and consider the \texttt{KB}-Hindi ($\approx 150 \ hours$), Tamil ($\approx 185 \ hours$) and Kannada ($\approx 65 \ hours$ train dataset as unlabeled audio sets. We set $\delta \geq 0.8$ for acquiring pseudo-labels in \texttt{TeLeS-A} experiment.

%For Active Learning experiments, we use the hindi \texttt{ASR} model trained on \texttt{PB} data as the base model. We choose Hindi Gramvaani dataset (reference add) as the target. We choose Gramvaani because the audio files are of telephone quality. The audio recordings are from crowd workers across India and, hence, we can find regional/dialectal speech variations. This dataset is considerably different from the dataset used for training the base \texttt{ASR} model, and hence, active annotations are imperative. The experiments in this section compare the efficiency of proposed acquisition score and that of the \texttt{SOTA} methods for active learning.
%The base \texttt{ASR} model is trained on limited data whose domain is entirely different from the target and hence, this experiment is chosen to assess the efficiency of proposed acquisition score and \texttt{SOTA} methods.

\textit{Baseline Approaches-} We compare with three baselines. The first approach uses the length-normalized path probability approach given in \cite{al2}\cite{al3}. The second approach is \texttt{SMCA}, presented in \cite{al1}. The final method is the \texttt{LMC} methodology proposed in \cite{al1}, where we build a 3-gram \texttt{KenLM} \cite{ken} using the transcripts to train the base \texttt{ASR} model and compute the perplexity score for acquiring data for annotation.

\subsection{Evaluation Metrics}

 We use the standard \texttt{WER (\%)} $\downarrow$ and \texttt{CER (\%)} $\downarrow$ metrics to evaluate the active learning approaches. We observe how much \texttt{WER} and \texttt{CER} reduce after finetuning the base \texttt{ASR} models with the acquired data. Each experiment is run for 30 epochs with same parameters used to train the base \texttt{ASR} models.
 
%To evaluate \texttt{TeLeS-A}-based active learning approach, we use the standard confusion matrix metrics such as accuracy, specificity, sensitivity,  positive predictive value, negative predictive value, and F1 score. The motive of active learning is to choose the set of audios that give worse transcripts when given to the base \texttt{ASR} model. The true positive bin has the count of audios that are chosen for manual annotation and its experimental \texttt{WER} (computed by comparing ground truth and prediction from \texttt{ASR} model) is greater than an arbitrary threshold. The true negative bin holds the count of audios with conditions converse to that of the true positive bin. 

%We arbitrarily fix \texttt{WER} threshold as 0.3. To fix \texttt{TeLeS-A} acquisition score thresholds, we do a systematic search of various thresholds in the range of [0, 1] as seen in Figure \ref{fig:qual} and compute the accuracy. We find that threshold values .., ..., ..., and ... are ideal for the four languages to . 

\subsection{Comparison with \texttt{SOTA}}

We fix the labelling budget based on audio hours. We choose the experimental label budget as a proportion $(1/10, 1/7)$ of the training dataset. 

Tables~\ref{al-hin}, \ref{al-tam}, and \ref{al-kan} report the \texttt{WER} and \texttt{CER} statistics on \texttt{KB} test-known sets. As seen in the statistics, \texttt{TeLeS-A} has better performance than \texttt{SOTA} methods. \texttt{LMC} depends on the \texttt{LM}, trained on the source \texttt{ASR} dataset and has no statistics about the target dataset transcripts. In a real-time scenario, there is no guarantee that obtaining transcripts from the target domain is possible. Thus, \texttt{LMC} is unreliable for real-time implementation, as evidenced by the observed \texttt{WER} in the tables. \texttt{SMCA} method performs better than \texttt{LMC} as it varies the dropout of the base \texttt{ASR} model and sees if the inference is similar even after model perturbation. For the \texttt{SMCA} experiment, we configure ten seeds to prepare ten different perturbed models, which may be a significant run-time overhead in cases of huge unlabeled datasets. However, the base \texttt{ASR} model's encoder weights are the same, and we cannot expect the perturbed model to expose uncertainty in all the circumstances. The \texttt{TeLeS-A} method is also better than the \texttt{SOTA} method as it acquires the set of confident samples, which aids in convergence.

%Figure .. shows the progress of reduction of error rate across epochs while acquiring 60\% of data from unlabeled training pool. Figure .. shows that usage of \texttt{TeLeS-A} acquired dataset convergence is better than \texttt{SOTA} methods.

%https://tex.stackexchange.com/questions/509171/how-to-highlight-a-point-in-line-graph-by-drawing-a-circle-around-it

\begin{table}[ht]
%\scriptsize
    \centering
   %\small
    \caption{Active Learning - \texttt{KB} Hindi Test-Known Dataset}
   \begin{tabular}{|p{1.1cm}|p{1.4cm}|p{1.4cm}|p{1.4cm}|p{1.4cm}|}
    \hline
        \textbf{Acquired Data Proportion} & \textbf{Path Probability (\texttt{WER/CER $\downarrow$})} & \textbf{SMCA (\texttt{WER/CER $\downarrow$})} & \textbf{LMC (\texttt{WER/CER $\downarrow$})} & \textbf{TeLeS (\texttt{WER/CER $\downarrow$})}\\
         \hline
         \texttt{$1/10$}  & 43.53/15.22 & 59.97/23.01 & 92.55/53.16  & \textbf{32.96/11.09} \\
         \hline
        \texttt{$1/7$} &  31.71/10.59 & 33.09/11.00 & 36.93/12.42  & \textbf{28.78/9.51} \\
         \hline
        
    \end{tabular}
    \label{al-hin}
\end{table}

\begin{table}[ht]
%\scriptsize
    \centering
   %\small
    \caption{Active Learning - \texttt{KB} Tamil Test-Known Dataset}
   \begin{tabular}{|p{1.1cm}|p{1.4cm}|p{1.4cm}|p{1.4cm}|p{1.4cm}|}
    \hline
        \textbf{Acquired Data Proportion} & \textbf{Path Probability (\texttt{WER/CER $\downarrow$})} & \textbf{SMCA (\texttt{WER/CER $\downarrow$})} & \textbf{LMC (\texttt{WER/CER $\downarrow$})} & \textbf{TeLeS (\texttt{WER/CER $\downarrow$})}\\
         \hline
         \texttt{$1/10$}  & 50.05/10.52 & 49.63/10.43 & 50.64/10.71  & \textbf{47.38/9.83} \\
         \hline
        \texttt{$1/7$} &  45.62/9.28 & 45.11/9.13 & 45.49/9.19  & \textbf{44.35/8.87} \\
         \hline
        
    \end{tabular}
    \label{al-tam}
\end{table}

\begin{table}[ht]
%\scriptsize
    \centering
   %\small
    \caption{Active Learning - \texttt{KB} Kannada Test-Known Dataset}
   \begin{tabular}{|p{1.1cm}|p{1.4cm}|p{1.4cm}|p{1.4cm}|p{1.4cm}|}
    \hline
        \textbf{Acquired Data Proportion} & \textbf{Path Probability (\texttt{WER/CER $\downarrow$})} & \textbf{SMCA (\texttt{WER/CER $\downarrow$})} & \textbf{LMC (\texttt{WER/CER $\downarrow$})} & \textbf{TeLeS (\texttt{WER/CER $\downarrow$})}\\
         \hline
         \texttt{$1/10$}  & 54.29/12.22 & 99.97/85.58 & 99.95/93.97  & \textbf{47.98/10.71} \\
         \hline
        \texttt{$1/7$} &  43.75/9.57 & 60.87/13.67 & 74.47/17.69  & \textbf{41.91/9.15} \\
         \hline
        
    \end{tabular}
    \label{al-kan}
\end{table}

\section{Conclusion and Future Work}

The manuscript examines the cons of class-probability-based confidence scores, entropy-based confidence scores, and \texttt{CEM} trained with binary target labels. We propose a better confidence target score that includes temporal and lexeme similarity. In addition to the score proposal, our work also gives a solution to handle the data imbalance issue in the confidence neural net training data. We test the efficacy of the score in three different language \texttt{ASR} models, each trained with varying hours of speech data. Through several evaluation metrics and graph-based investigations, we find that the proposed score is better than the \texttt{SOTA} approaches. The usability of \texttt{TeLeS-WLC} in practical \texttt{ASR} downstream active learning task further illustrates that \texttt{TeLeS} is a reliable score. Current work can be extended to estimate confidence of deletion of words in the utterances. Our next plan is to design an appropriate module to detect the deletions. Also, one could experiment with the usability of the proposed method for other downstream and upstream tasks. Future research direction could be to adapt a model trained in one language to another language.

\section*{Acknowledgments}
Prasar Bharati and MeitY, India funded the project. We acknowledge the assistance of the project’s stakeholders for their consistent support and Param Siddhi for providing compute infrastructure. Contribution of IIT Kanpur student interns Raghav Karan and Mridul Pandey towards the project are noteworthy. We also recognise the data annotation team members- Divyanshu Tripathi, Anika Kumari, Ankita Bhattacharya, Shruthi Mishra and Sachindananda Prajapathi for their meticulous effort in preparing the data.

%\newpage

%\vskip 0pt plus -1fil

\begin{IEEEbiography}[{\includegraphics[width=1in,height=1.25in,clip,keepaspectratio]{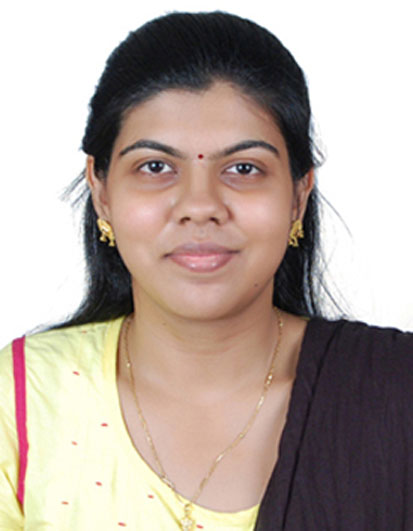}}]{Nagarathna Ravi}
received the B.Tech. degree in information technology from the Madras Institute of Technology, Anna University, India, and the M.E. degree in computer science engineering from the Thiagarajar College of Engineering, India. She received the Ph.D. degree from Anna University, India. She is currently a Postdoctoral Researcher at the Indian Institute of Technology (IIT) Kanpur, India. Her research interests are ASR, SDN and IoT.
\end{IEEEbiography}

\vskip 0pt plus -1fil

\begin{IEEEbiography}[{\includegraphics[width=1in,height=1.25in,clip,keepaspectratio]{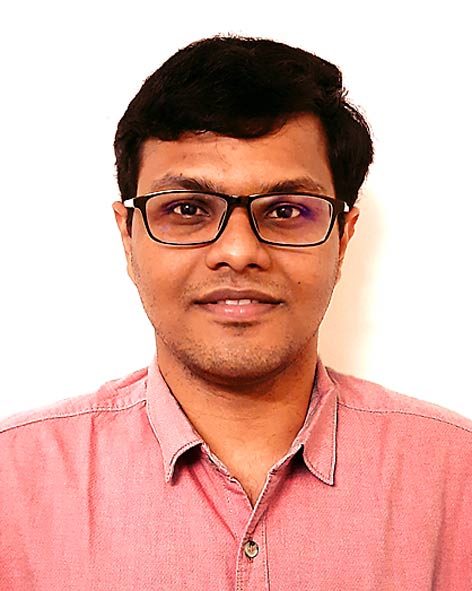}}]{Thishyan Raj T}
received the B.E. degree in Electronics and Communications Engineering from the PESIT Bangalore South Campus. He is currently pursuing M.S. degree in Department of Electrical Engineering, IIT Kanpur. His research interests are ASR, audio processing, and machine learning.
\end{IEEEbiography}

\vskip 0pt plus -1fil

\begin{IEEEbiography}[{\includegraphics[width=1in,height=1.25in,clip,keepaspectratio]{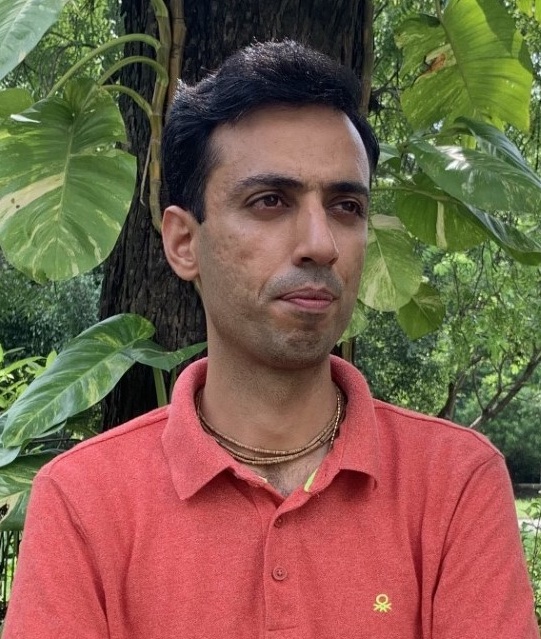}}]{Vipul Arora}
received the B.Tech. and Ph.D. degrees in electrical engineering from the Indian Institute of Technology (IIT) Kanpur, India, in
2009 and 2015, respectively. He has been a Postdoctoral Researcher at the University of Oxford and a Research Scientist at Amazon
Alexa, Boston, MA, USA. He is currently working as an Associate Professor with the Department of Electrical Engineering, IIT Kanpur. His research interests include machine learning, audio processing, machine learning for physics, and time series analysis.
\end{IEEEbiography}

\end{document}